\documentclass{aa}  

\usepackage{graphicx}
\usepackage{txfonts}
\usepackage[english]{babel}
\usepackage[colorlinks=true, allcolors=blue]{hyperref}
\begin{document}

 \title{Testing the extended corona model with the optical/UV reverberation mapping of the accretion disk}
 \titlerunning{Extended corona model}


   \author{Vikram K. Jaiswal
          \inst{1}
          \and
         Bo\. zena Czerny
         \inst{1}
          }

   \institute{Center for Theoretical Physics, Polish Academy of Sciences, Al. Lotnik\' ow 32/46, 02-668 Warsaw, Poland}

   \date{Received ....}

 
  \abstract
   {The illumination of the accretion disks is frequently studied assuming that the incident X-ray flux is a point-like source. The approach is referred as lamppost model.}
   {The most recent computations of the X-ray reprocessing by the disk take into account the departure from the simple lamppost models. However, in computations of the incident flux thermalization and subsequent re-emission in the optical-UV band the lamppost approximation is most frequently assumed. We test if the UV-optical reverberation mapping and time delay measurements are sensitive to this assumption.}
   {We assume that the incident radiation originates from a region extended along the symmetry axis. To model this, we adopt a simple setup by representing the emission as two lamps irradiating the disk simultaneously from two different heights. We then compare the resulting predictions with those obtained for a single lamppost located at an intermediate height.}
   {We show at the basis of the transfer function that the deviation of the wavelength-dependent delay curve shows at most a difference of 20 \% in comparison to a single lamppost, assuming the black hole mass of $10^8 M_{\odot}$, Eddington ratio 1, and the location of the lamps at 5 and 100 r$g$. The maximum deviation happens for the lamp luminosity ratio $\sim3$. When simulating light curves for a two-lamp setup and a standard lamppost with the same black hole mass and a sampling rate of 0.1 days, we find no measurable differences in the ICCF profiles between the two setups. Larger black hole mass and considerably lower Eddington ratio would allow to see larger differences between a single lamppost and a two-lampost model.}
   {UV/optical reverberation mapping is not very sensitive to the vertical extension of the corona. To detect such extension we would need small errors in the time delay (below 20 \% in optimized setup, and much less otherwise), and the sampling of the lightcurves must be very dense.  The effective position of the lamp is given by the luminosity-weighted position of the emitting components. }


   \maketitle
%

\section{Introduction}

Active Galactic Nuclei (AGN) are highly variable systems powered by accretion onto supermassive black holes. Most of the energy is released in the innermost regions of the accretion flow, deep within the gravitational potential well of the central black hole. This complex and dynamic region undergoes rapid variations \citep[see, e.g.,][]{krolik_book1999}. The UV and X-ray emission generated in these inner regions irradiates the more distant parts of the accretion flow, which, in efficiently accreting systems, typically takes the form of a geometrically thin, optically thick accretion disk.

The production of the X-ray continuum involves seed photons from the accretion disk being Comptonized by high-energy electrons in the corona. This interaction up-scatters the photons to higher energies, resulting in an X-ray spectrum. A portion of these X-ray photons interacts with the accretion disk, illuminating it, while the rest travel unimpeded directly to the observer. The phenomenon of X-ray irradiation of the accretion disk has been extensively studied in AGN over the years \citep[e.g.,][]{nandra1989, pounds1990, george_fabian1991}.
This irradiation leads to the formation of characteristic X-ray features, such as the iron K$\alpha$ line and the Compton hump. These features originate in the accretion disk close to the black hole, where relativistic effects become significant \citep{iwasawa1996, fabian2000, dovciak2004}.

The first measurement of a delay in reprocessed X-rays, known as a reverberation lag, was made in the source Ark 564 using XMM-Newton observations, as reported by \citet{Mchardy2007}. Subsequently, \citet{Fabian2009} measured the lag between the soft X-ray energy band (0.3–1 keV) and the hard band (1–4 keV) in the AGN 1H 0707$-$495. Since then, numerous reverberation lags have been detected in AGN \citep{Emman2011, kara2013, zoghbi2013, cackett2013, cackett2014, kara2016b,kara2016a, ursini2020}. For a detailed review, see \citet{deMarco2019}, and for a broader discussion of light echo studies, refer to \citet{Cacket_Sci2021}.

More distant regions of the accretion disk are also illuminated by X-rays and UV radiation. In these regions, most of the radiation is absorbed, temporarily increasing the local disk temperature. This process results in observable time delays between subsequent optical bands due to the light travel time effect. Early studies of the optical/UV continuum focused on measuring inter-band time delays \citep[e.g.,][]{collier1998, collier1999, oknyanskij2003, Sergeev2005, cackett2007, poindexter2008, fausnaugh2016, Montano2022}.
More recently, reverberation mapping campaigns utilizing HST, SWIFT, and ground-based telescopes have extended the spectral coverage to include the X-ray and far-UV regions alongside optical bands \citep{McHardy2014, McHardy2018, cackett2018, cackett2020, vincentelli2021, cackett2023, edelson2024, liu2024}. This multi-band, broad-wavelength photometric approach provides a powerful tool for independently probing the geometry of the hot corona and the inner accretion flow.

Despite extensive observations and studies, the exact geometry of the corona remains poorly understood. Historically, three primary models have been proposed to describe the hot corona:
\begin{itemize}
\item     Flat Disk Atmosphere Model: This model describes the corona as a hot, thin atmosphere that lies above the flat, geometrically thin accretion disk. Soft UV photons emitted vertically from the disk interact with the corona's relativistic electrons, where they are upscattered via inverse Compton scattering, producing the observed hard X-rays.  \citep{Haardt1993, Haardt1994, galeev1979}.
\item    Hot, Thick Accretion Flow Model: Here, the corona is envisioned as a hot, optically thin accretion flow surrounding the black hole \citep{ichimaru1977,Narayan1994,Esin1997}.
\item    Lamppost Model: This model posits that the corona is a compact source located along the black hole's spin axis, often approximated as a point source at a height $h$ above the disk \citep{Matt1991, rokaki1993, Beloborodov1999}. It may also be associated with the base of a relativistic jet  \citep{henri1991, Markoff2005}.
    \end{itemize}

Advances in spectral modeling have shown that the first scenario cannot produce the observed hard X-ray spectral slopes, and the relative importance of the second and third models appears to correlate with the source's Eddington ratio \citep[see, e.g.,][]{rozanska2015, giustini2019, ballantyne2024, palit2024}.
Recent theoretical work, such as the broad-band spectral energy distribution (SED) models by \citet{kubota2018}, integrates contributions from the hot corona, a warm corona, and an outer cold disk. These models address the energy budget but do not resolve the geometric details of the corona.

In this study, we focus on sources radiating above a few percent of the Eddington luminosity, where the inner hot flow is unlikely to dominate. For such sources, the lamppost geometry provides a plausible approximation. While the Comptonization process requires the corona to have an extended structure to effectively intercept disk photons, the lamppost model remains a reasonable simplification if radial extent of corona is small compared to its height. Nevertheless, this approximation needs to be rigorously tested against observational data.


In recent years, researchers have increasingly explored the possibility of an extended hot corona using observational data. Several X-ray reprocessing studies have incorporated the effects of an extended, contracting, or moving corona \citep{wilkins2012, wilkins2016, kara2019, Chainakun_2019, You2021}.
\citet{Szanecki2020} introduced a new extended lamppost model that considers the spatial extent and rotational motion of the X-ray source, applying it to the AGN 1H 0707$-$495. However, their study did not yield definitive evidence for the size of the corona. Building on this, \citet{Hancock2023} developed a two-corona model for the sources 1H 0707$-$495 and IRAS 13224$-$3809, suggesting that the corona could extend up to approximately 20  $r_g$ for IRAS 13224$-$3809, where rg refers gravitational radius.. Similarly, \citet{Lucchini2023} investigated the same source and demonstrated that the two-corona model can successfully reproduce its spectral and timing properties.
Additionally, \citet{ursini_two_corona2020} proposed a complex structure to explain both the hard and soft X-ray emission in the AGN HE 1143-1810. Their model incorporates two components: a hot, hard X-ray-emitting corona and a warm corona, consistent with the multi-component framework discussed earlier.

These findings underscore the growing interest in refining corona models to better understand their spatial and physical properties. The issue of the corona height also 
 plays a crucial role in modeling the variable X-ray radiation incident on and absorbed by the outer regions of the accretion disk. This irradiation drives UV and optical variability, which serves as a key tool for probing the structure of the cold outer accretion disk and the broad line region (BLR). Such as, the reverberation mapping technique, has been extensively utilized in numerous studies, primarily to measure black hole masses by establishing the radius-luminosity relationship for strong emission lines \citep{peterson88, peterson1993, peterson2004, bentz2013, zajacek2021, shen2024}.

While the height of the corona is not a significant factor for BLR mapping, it has emerged as a critical parameter in recent accretion disk reverberation mapping models \citep{kammoun2021, kammoun2023}. Representative values for the corona height have been derived through data fitting in these studies \citep{kammoun_data2021}, highlighting its importance in understanding disk irradiation and variability.

One of the most intriguing findings from these two lines of research is the apparent discrepancy in the inferred heights of the corona across X-ray and optical/UV delay studies. In X-ray studies, the corona height is typically small, varying from 5 $r_g$ to 39 $r_g$. In contrast, optical/UV studies suggest a significantly larger height, varying from 22 $r_g$ to 74 $r_g$

This disparity raises important questions about the nature of the X-ray source. It is unclear whether the position of the X-ray source genuinely varies between the studied objects or if the corona is vertically extended, causing the inner and outer regions of the disk to respond to irradiation from different parts of this structure. Further investigation is essential to reconcile these findings and to understand the complex interplay between the corona's geometry and the multiwavelength variability it induces.

This paper aims to explore the potential impact of a vertically extended corona on UV/optical reverberation mapping. To model this scenario, we approximate the corona as two distinct lamppost-like sources positioned along the symmetry axis, rather than a single point source. Using this setup, we simulate the irradiation of the accretion disk to investigate how the extended corona influences the observed variability.
Our objective is to assess whether optical/UV data can effectively differentiate between a single lamppost corona and a vertically extended corona, providing new insights into the geometry of the X-ray emitting region.

\section{Method}

\begin{figure}
\centering
\includegraphics[width=0.50\textwidth]{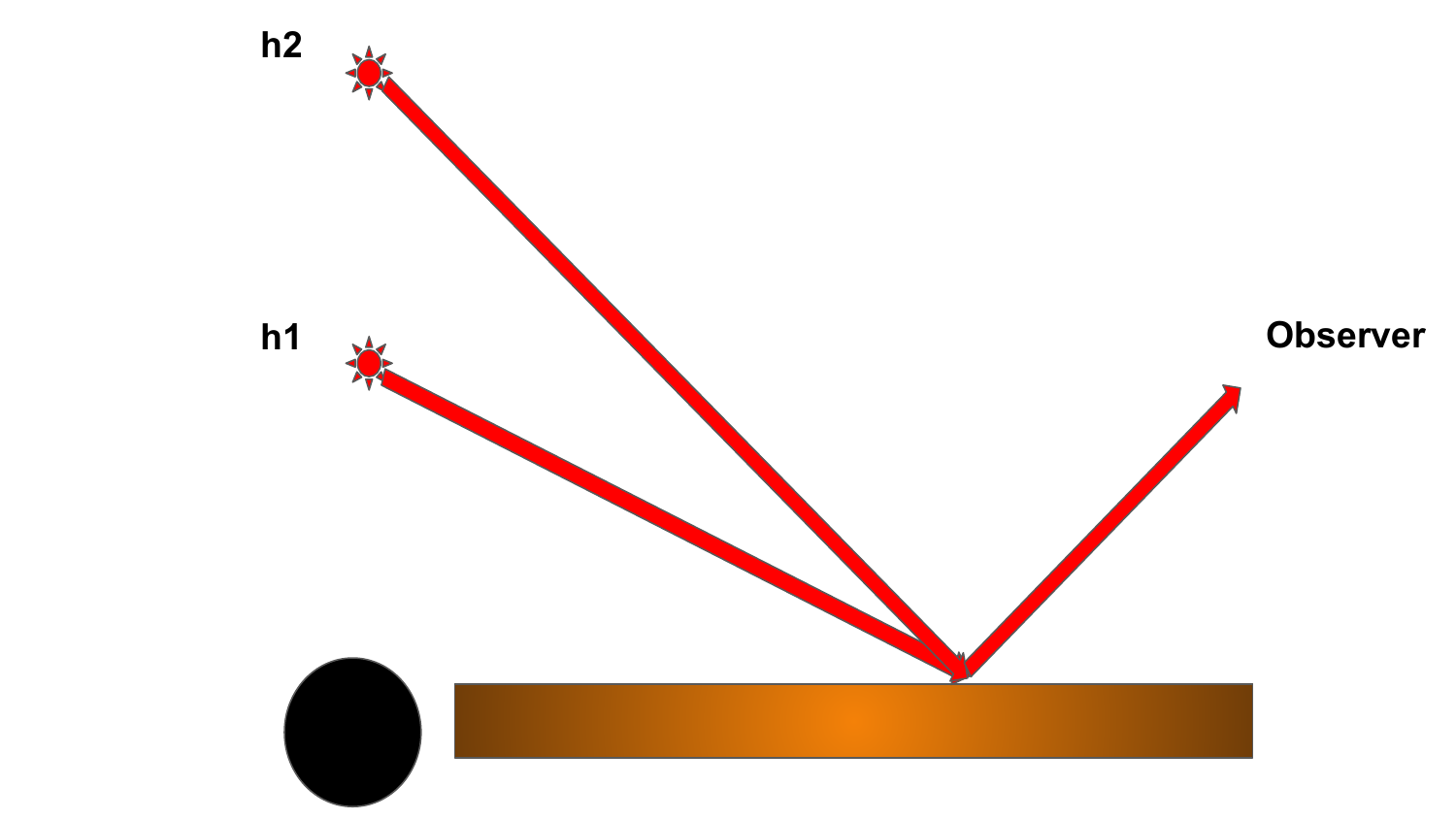}
\includegraphics[width=0.45\textwidth]{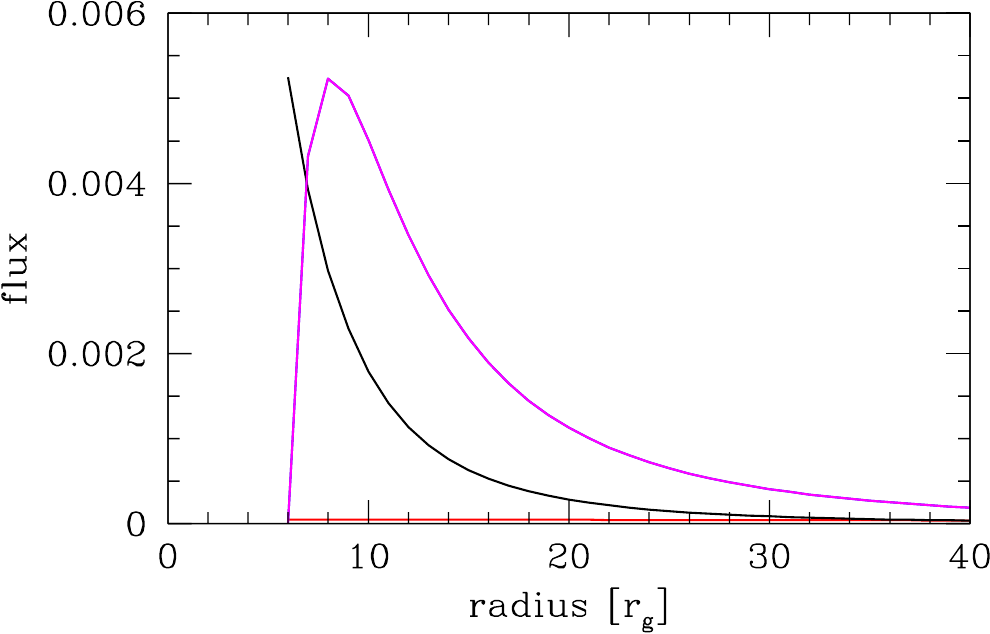}
\includegraphics[width=0.45\textwidth]{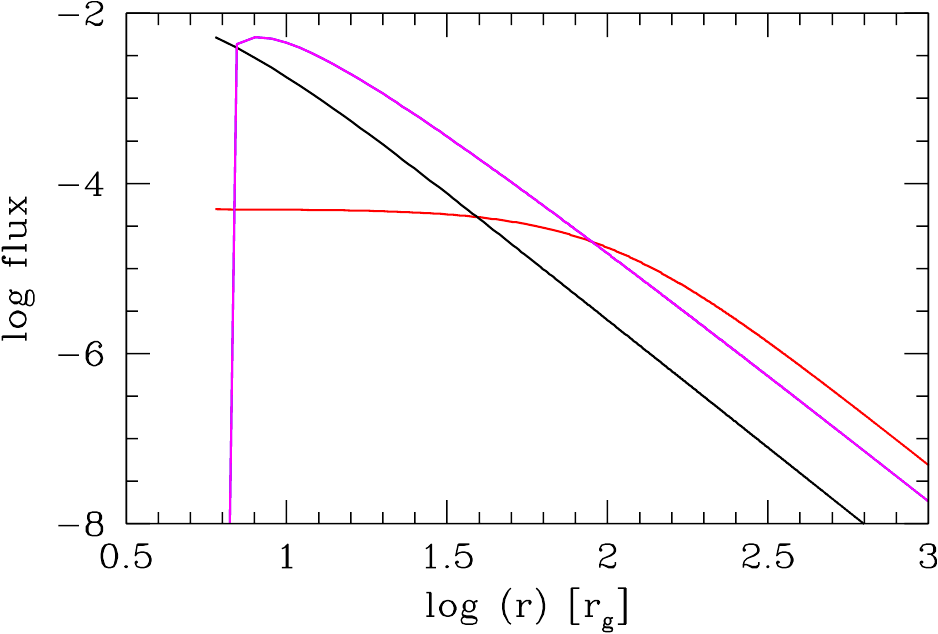}
\caption{A schematic representation of the geometry is shown in the upper panel, along with an example of the stationary incident flux for this geometrical setup presented in linear scale (middle panel) and logarithmic scale (lower panel). The corona heights are 5 $r_g$ (black line) and 100 $r_g$ (red line), each contributing approximately 30\% of the total disk luminosity. The disk flux is represented in magenta.
}
\label{fig:schematic}
\end{figure}

We simulate the reprocessing of the coronal emission extended along the symmetry axis in a very simple way. We assume that instead of a single point-like source located at the symmetry axis we have two point-like sources located at a different heights, $h_1$ and $h_2$. The average bolometric luminosities of the two components are assumed to be $L_1$, and $L_2$, and in general they can be different. The setup is illustrated in Figure~\ref{fig:schematic}.

In addition, we illustrate the stationary incident flux as a function of the radius in the case of two coronas of the same luminosity, and compare it with the disk flux from dissipation. At large distances all fluxes are proportional to $r^{-3}$ but at small radii the distributions are quite different. As shown in middle and bottom panel of Figure~\ref{fig:schematic}, in the case of small height the irradiation dominates close to ISCO, and before the peak of the disk emission drops below and remains proportional to the disk flux. For larger height the incident flux is flat at to the radius of the order of the lamp height, well below the disk emission, and then starts to dominate by a small factor, decreasing with the radius in proportion to the disk flux. 

\subsection{reprocessing of the radiation by the accretion disk}
\label{sect:disk}

We use a simple accretion disk model of \citet{SS1973} for this purpose, and geometrical optics for light propagation. It is a reasonable approximation for sources with Eddington rate above a few percent. In such sources the cold optically thick geometrically thin disk extends down to innermost stable circular orbit (ISCO). In Shakura-Sunyaev model ISCO is located at 6 $r_g$ (where $r_g = G M_{BH}/c^2)$, and $M_{BH}$ is the black hole mass, while G and c are gravitational constant and the light velocity). We neglect all general relativity effects, which is a good approximation for a non-rotating black hole. We concentrate on the relative effects of having two instead of one lampost sources, and our simplified treatment shows the direction of the change.

The disk is divided into zones as in \citet{jaiswal2023}.  In order to calculate the effect of incident radiation, in 
the first step we set a quasi-logarithmic radial grid that covers the accretion disk between $R_{isco}$ and $R_{out}$, with a variable density to ensure proper resolution at the inner parts. Specifically, we define the radial bin size using the formula: $dR = 0.085*(\frac{R}{R_{\rm isco}})^{0.85}$. For each value of "R," we also increase the grid step in the angular ($\phi$) direction by $d\phi = \frac{1.5700}{N_{div}}$, where $N_{div}=1000$.  Once we have a given "R" and $\phi$ value, we calculate the (x,y) Cartesian coordinate in the disk plane and the surface element, $ds = R*dR*d\phi$ in a unit of $Rg^2$. We neglect the disk's height and assume $z = 0$.
For a given (x,y) coordinate, we calculate the total delay $\tau_{total}(x,y)$, which is the sum of the time $\tau_d(r)$ taken by a photon to reach a given disk location from the bottom corona located at height $h_1$ along the symmetry axis to the accretion disk and the time taken to reach the observer after reprocessing by the disk, $\tau_{do}(x,y)$. We define this time delay with respect to the plane crossing the equatorial plane at $R_{out}$, $\phi=0$, and perpendicular to the direction towards the observer. This delay thus depends on the inclination angle $i$, the corona height, the position on the accretion disk, and the black hole mass. Similarly, we calculate the delay for the top corona located at $h_2$.

The disk temperature is calculated from the sum of the flux generated by the disk (parametrized by $M_{BH}$ and accretion rate, $\dot M$, or, equivalently, the Eddington rate) and the incident flux. We assume black body local radiation, not applying any color-corrections.

We are assuming that the incident radiation is perfectly thermally absorbed. Whenever an X-ray photon strikes the disk, all the radiation it carries is absorbed by the disk, causing a local increase in the disk temperature. Therefore, in our model, we have ignored any energy-dependent reflection, as discussed in \citet{Kammoun2019} and \citet{kammoun2021}, and have instead considered that all the incident radiation is absorbed by the disk and then reprocessed, i.e. considering perfect thermalization of incident X-ray flux. 

\subsection{Transfer function}

The computation of the transfer function involves using a short impulse of light and determining the corresponding reprocessed flux as a function of time. The exact width of the pulse is not critical; a typical value is 0.05 days. For a black hole mass of
$10^8 M_{\odot}$, often adopted as a reference, this pulse duration corresponds to a light travel time of $1.76 r_g$, which is well within the $10 r_g$
 limit recommended by \citet{kammoun2021}. To calculate the response function for two lamps, a pulse is sent from each corona for a duration of 0.05 days. At any given radius (r) and angle ($\phi$), there are two possible scenarios: the pulses either overlap or they do not. If the pulses do not overlap, we calculate the total flux separately, as the two pulses will arrive independently. However, when the pulses do overlap, we first calculate the flux from corona 1, then we compute the combined flux from both coronas, and finally, we calculate the flux from corona-2. This method allows us to generate the response function.


\subsection{Creation of corona and reprocessed lightcurves}
\label{sect:creating_curves}

The use of the response function requires that the variability be of small amplitude, ensuring that a single response function can describe all the disk luminosity states. In numerical method we do not require that the variable irradiating flux creates only a small amplitude variations. We generate the variable flux in some timestep bins, and we calculated the reprocessing of the incident radiation in each disk element and time bin independently. This is a much more time-consuming method but it reflects well the reality, including all the sampling issues.

We start with creating a single long equal-step lightcurve representing the corona using the \citet{timmer1995} algorithm which is parameterized by three slopes and two frequency breaks. Such description is more general than damped random walk model applied to AGN by \citet{kelly2009} which is equivalent to the fixed slopes, 0 and -2, and a single break. 
The random curve is additionally parameterized by the total variance which sets the fractional amplitude of the variability. The corona light curve is produced by multiplying the fixed luminosity with the dimensionless random number series, and the resulting series is utilized as the irradiation source for the accretion disk. In the next step we apply the observationally motivated sampling of the dense lightcurve. In this paper we assumed 0.1 day sampling, and the duration of the lightcurve 200 days.

\begin{figure}
\centering
\includegraphics[width=0.6\textwidth]{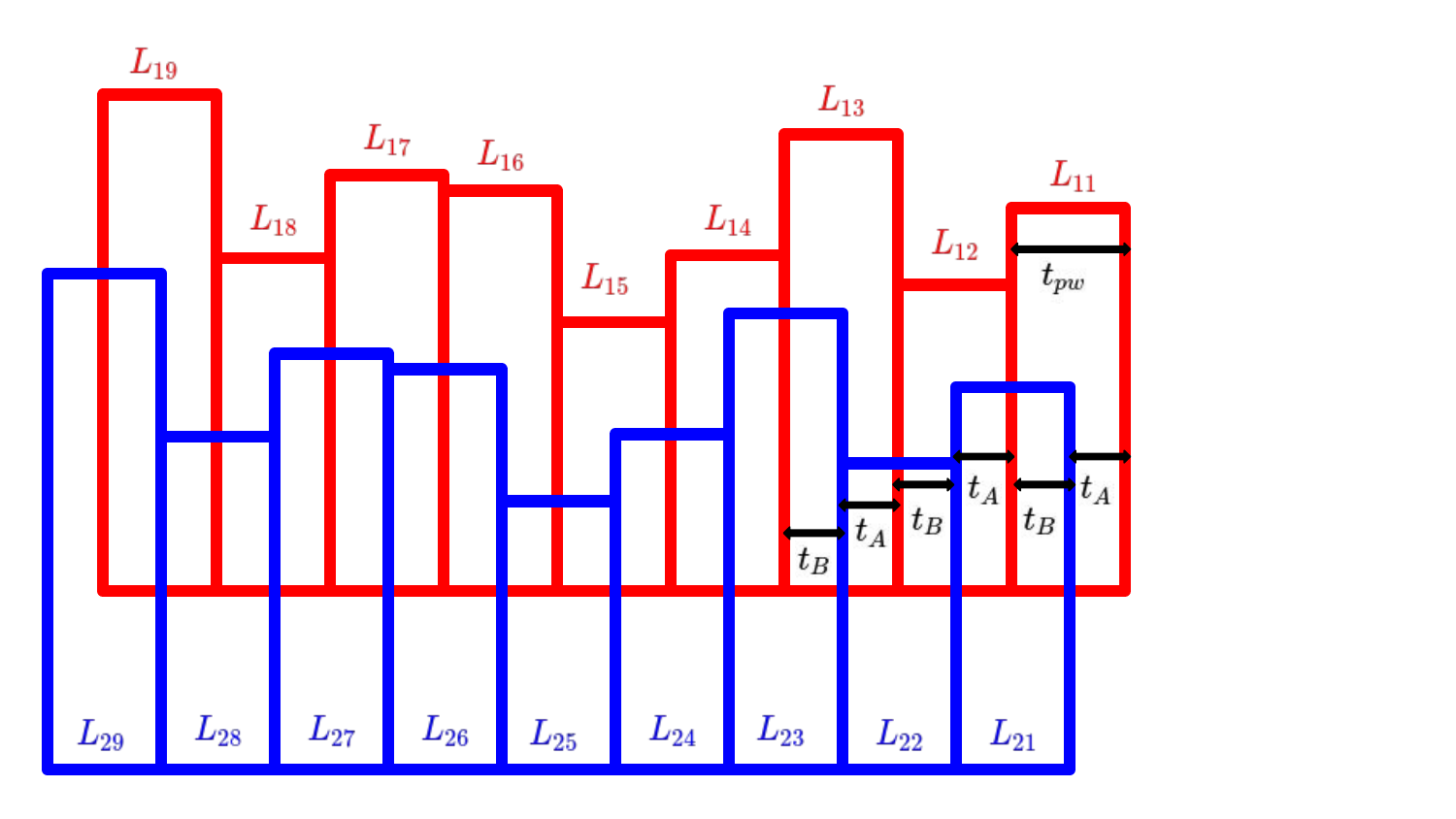}

\caption{ The diagram illustrates the light curves from Corona-1 (red) and Corona-2 (blue) at a specific location on the disk. Here, $t_{pw}$ represents the pulse width of the light curves, $t_A$ denotes the temporal shift between the two for a given $(r, \phi)$, which varies based on the disk location, and $t_C$ is the result of subtracting the shift ($t_A$) from the pulse width ($t_{pw}$).}

 \label{fig:2c_light_curve}
\end{figure}

In the case of two coronal sources we usually use the same dense curve, without any intrinsic delay between the coronal points. Next we calculate the reprocessing of the incident radiation at every timestep. As illustrated in Figure~\ref{fig:2c_light_curve}, the accretion disk receives radiation at any radius \( r \) and azimuthal angle \( \phi \) from two distinct radiation sources, referred to as corona1 and corona2. These sources emit light that reaches different locations on the disk at different times due to their varying spatial positions and distances from the disk. For a specific point on the disk, \((r, \phi)\), the light pulse from corona1 arrives at time \( t_0 \), while the light pulse from corona2 reaches the same location later, at time \( t_0 + t_A \). Here, \( t_A \) represents the additional time delay caused by the difference in light travel paths from corona2 to the disk location.
To simplify calculations, the red pulse ('\( L_{11} \)') emitted by corona1 is ignored during the initial phase because the blue pulse ('\( L_{21} \)') from corona2 is absent in this period. The analysis begins at time \( t_0 + t_{pw }\), where \( t_{pw} \) pulse width. During the time interval from \( t_0 + t_{pw} \) to \( t_0 + t_{pw} + t_A \), the total flux \( F_\text{irr} \) and the corresponding effective temperature \( T_\text{irr} \) at the location \((r, \phi)\) are computed as:
\begin{equation}
F_\text{irr}(r, \phi) = \frac{3GM\Dot{M}}{8 \pi r^3}\left(1-\sqrt{\frac{6}{r}}\right) + \frac{L_{12}h_1}{4\pi r_1^3} + \frac{L_{21}h_2}{4\pi r_2^3}
\label{eq:eq1}
\end{equation}
\begin{equation}
T_\text{irr}(r, \phi) = \left[\frac{3GM\Dot{M}}{8 \pi r^3 \sigma_B}\left(1-\sqrt{\frac{6}{r}}\right) + \frac{L_{12}h_1}{4\pi r_1^3 \sigma_B} + \frac{L_{21}h_2}{4\pi r_2^3 \sigma_B}\right]^{\frac{1}{4}}
\label{eq:eq2}
\end{equation}
where \( r_1=\sqrt{r^2+h_1^2} \), \( r_2=\sqrt{r^2+h_2^2} \), \( L_{12} \)  and \( L_{21} \) are luminosities associated with the two sources, and \( h_1 \) and \( h_2 \) represent their respective heights above the disk.
For the subsequent time interval from \( t_0 + t_{pw} + t_A \) to \( t_0 + t_{pw} + t_A + t_B \), the conditions change as the sources evolve. During this phase, the flux and temperature are updated to:
\begin{equation}
F_\text{irr}(r, \phi) = \frac{3GM\Dot{M}}{8 \pi r^3}\left(1-\sqrt{\frac{6}{r}}\right) + \frac{L_{12}h_1}{4\pi r_1^3} + \frac{L_{22}h_2}{4\pi r_2^3}
\end{equation}
\label{eq:eq3}
\begin{equation}
T_\text{irr}(r, \phi) = \left[\frac{3GM\Dot{M}}{8 \pi r^3 \sigma_B}\left(1-\sqrt{\frac{6}{r}}\right) + \frac{L_{12}h_1}{4\pi r_1^3 \sigma_B} + \frac{L_{22}h_2}{4\pi r_2^3 \sigma_B}\right]^{\frac{1}{4}}
\label{eq:eq4}
\end{equation}
Once the effective temperature \( T_\text{irr} \) is determined, it becomes possible to compute the emitted intensity across wavelengths using the Planck function. This is expressed as:
\begin{equation}
I_\text{irr}(\lambda) = \frac{2 \pi h c^2}{\lambda^5} \frac{1}{e^{\frac{hc}{\lambda k_B T_\text{irr}}} - 1}
\label{eq:eq5}
\end{equation}
For a non-irradiated disk (without external illumination from the coronal sources), the flux \( F_\text{disk} \), effective temperature \( T_\text{disk} \), and intensity \( I_\text{disk}(\lambda) \) are given by:
\begin{equation}
F_\text{disk}(r, \phi) = \frac{3GM\Dot{M}}{8 \pi r^3}\left(1-\sqrt{\frac{6}{r}}\right)
\label{eq:eq6}
\end{equation}
\begin{equation}
T_\text{disk}(r, \phi) = \left[\frac{3GM\Dot{M}}{8 \pi r^3 \sigma_B}\left(1-\sqrt{\frac{6}{r}}\right)\right]^{\frac{1}{4}}
\label{eq:eq7}
\end{equation}
\begin{equation}
I_\text{disk}(\lambda) = \frac{2 \pi h c^2}{\lambda^5} \frac{1}{e^{\frac{hc}{\lambda k_B T_\text{disk}}} - 1}
\label{eq:eq8}
\end{equation}
The impact of the external illumination by the coronae can be studied by subtracting the contribution of the intrinsic disk emission from the total irradiated intensity. This yields the net effective intensity as:
\begin{equation}
I_\text{eff}(\lambda) = I_\text{irr}(\lambda) - I_\text{disk}(\lambda)
\label{eq:eq9}
\end{equation}
Finally, the monochromatic luminosity \( L_{\lambda} \) can be computed by multiplying the net effective intensity \( I_\text{eff}(\lambda) \) by the corresponding surface area of the emitting region:
\begin{equation}
L_{\lambda} = \int I_\text{eff}(\lambda) \, dA
\label{eq:eq10}
\end{equation}
This approach allows us to analyze how the external radiation sources contribute to the observed flux, temperature, and spectral properties of the disk, enabling a deeper understanding of the physical processes at play.


\section{Results}

We first analysed the response function from the extended
corona model and compare the resulting time delay pattern to see if this geometry can be distinguish from a simple lamppost model in the data. 

\subsection{response function from the extended corona}

\begin{figure}
\centering
\includegraphics[width=0.5\textwidth]{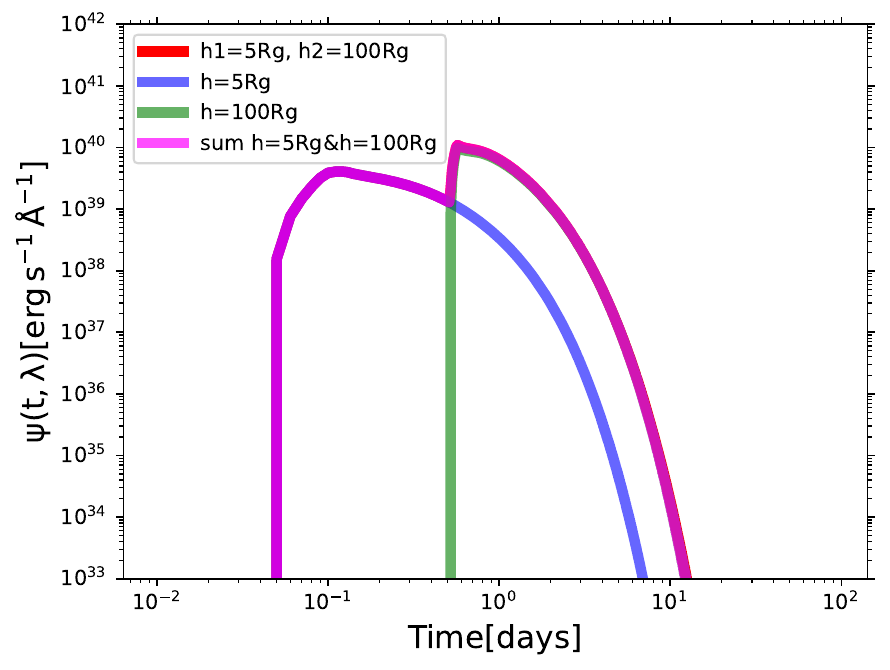}
\includegraphics[width=0.5\textwidth]{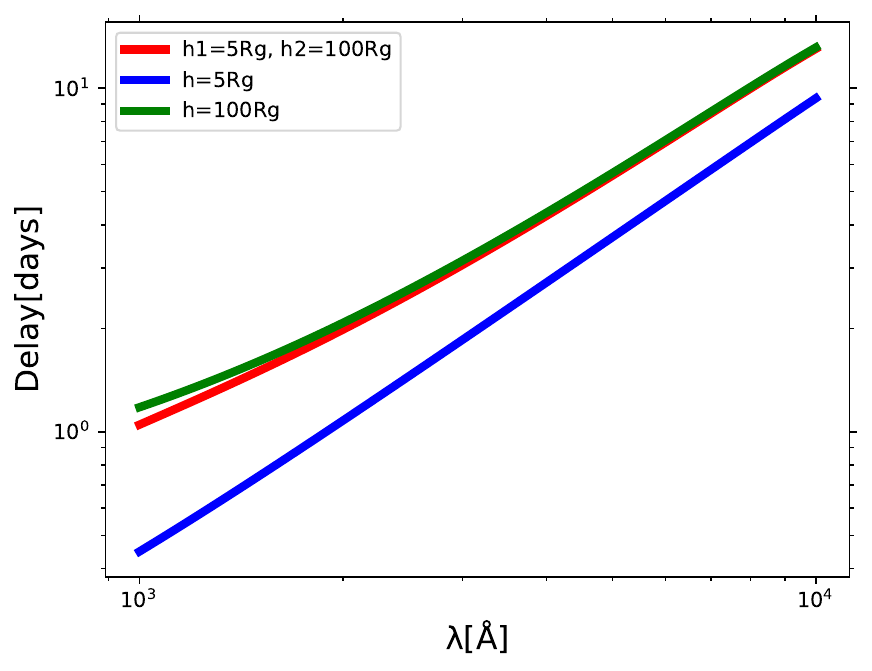}
\caption{The upper plot shows the response functions for a single corona with a height of $5r_g$ (blue) and $100r_g$ (green), as well as for two corona (red). The summed response function for the single corona at $5r_g$ and $100r_g$ is shown in pink. The combined response function closely overlaps with the two-corona response function for the given parameters. The bottom plot displays the corresponding delay plot. For the single corona, the parameters are: black hole mass $10^{8}M_{\odot}$, incident luminosity $L_X = 1.0 \times 10^{46}$ erg s$^{-1}$, Eddington ratio $= 1.0$, heights $h = 5r_g$ (blue) and $h = 100r_g$ (green), and viewing angle $i = 30^\circ$. For the two coronae, the parameters are: black hole mass $10^{8}M_{\odot}$, luminosities $L_1$ and $L_2 = 1.0 \times 10^{46}$ erg s$^{-1}$, Eddington ratio $= 1.0$, heights $h_1 = 5r_g$ and $h_2 = 100r_g$, and viewing angle $i = 30^\circ$.
}

 \label{fig:response1}
\end{figure}

 We have computed the response function for the two-lamppost corona model. An example for the wavelength $\lambda = 1000$ \AA~ is presented in Figure~\ref{fig:response1}(top), which illustrates four response functions, each distinguished by a unique color. The blue curve represents a single corona positioned at a height of $5 \, R_g$, while the green curve corresponds to a single corona at a height of $100 \, R_g$. The response for the two-lamp model is depicted in red. Both the combined response function and the two-point corona response function begin at the same time. Each single-corona response function exhibits a distinct peak that depends on the corona's height. However, the two-lamppost model produces a broader, two-peak response function, reflecting the travel time of light from both lamps to the disk and subsequently to the observer. For the given parameters, the two-lamp corona's response function can be approximated by summing the red and blue single-corona response functions (indicated in pink). Notably, the shape of the time delay curve in the two-corona case generally aligns with that of the single-corona case.
.

 

 \subsection{time delays from the extended corona}

Having the response functions for a range of wavelengths, we determined the the time delay for those three cases: two single lamp models and for a two-lamp model. The delay profiles are displayed in the bottom plot of figure ~\ref{fig:response1}. Single lamp solutions are equivalent to those  created in \citet{jaiswal2023}. They are consistent with the standard $\lambda^{4/3}$ trend when the corona height is small, and show systematic departure with the rise of the corona height \citep{kammoun2021}. The delay from the two-lamp model is roughly following the same trend. We do not see any sub-structure related to the two-component character of the transfer function. There is a small departure in the overall shape.

If we analyse the actual data, we do not know the lamp position. Therefore, in order to find a criterion for the presence of two lamps instead of one we must compare the two-lamp setup with the option of a single lamp setup, most similar in properties. To ensure accurate comparison, we adjusted the height of the single corona so that it matched the delay of the 2-corona for the longest wavelength, and measured the departure at the shortest wavelengths. We could not simply measure the departure from the $\lambda^{4/3}$ trend since the curvature effect is always present.

\begin{figure*}[ht!]
            \includegraphics[width=.3\textwidth]{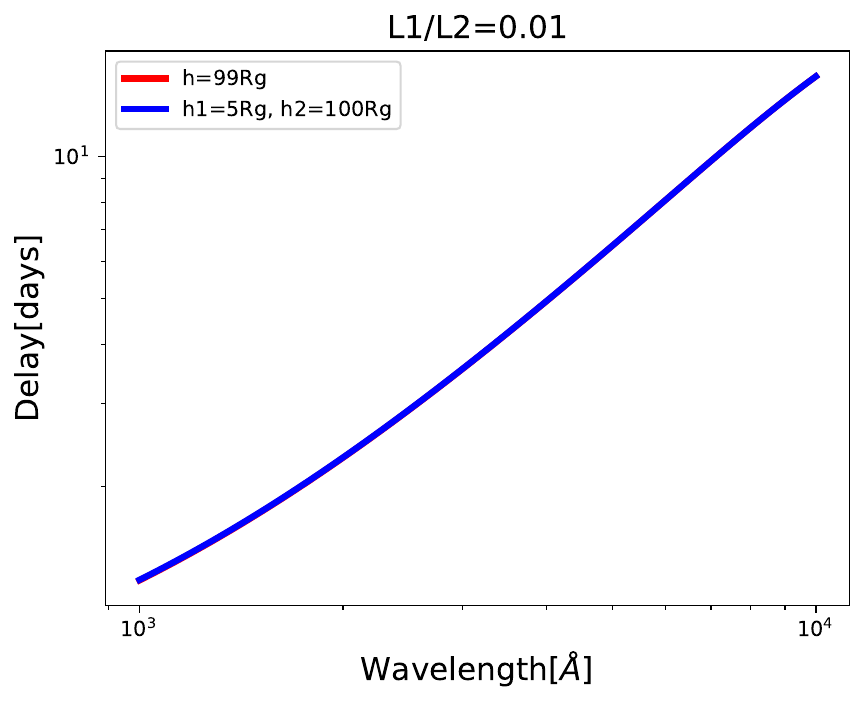}\hfill
            \includegraphics[width=.3\textwidth]{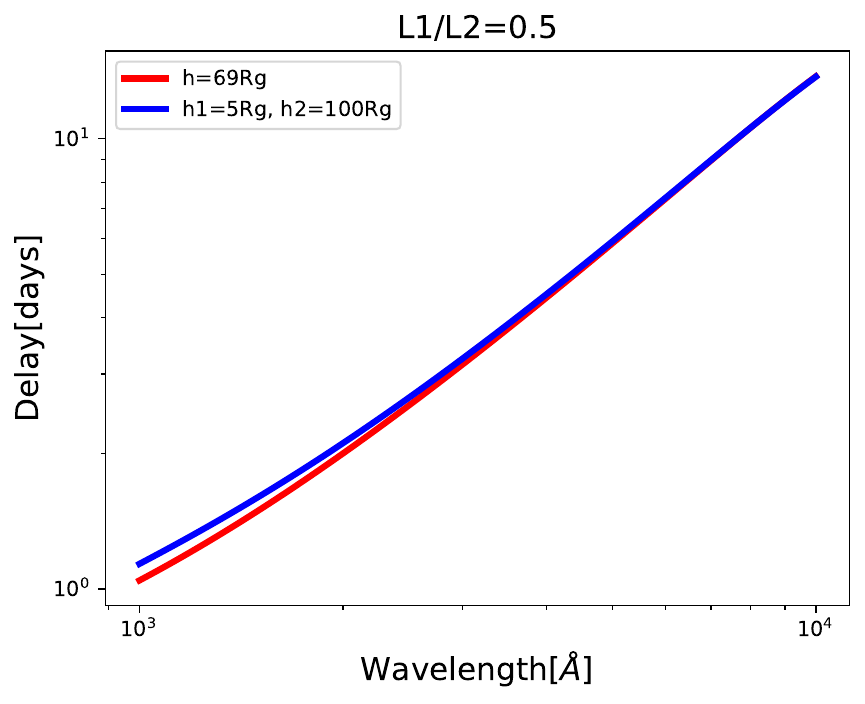}\hfill
            \includegraphics[width=.3\textwidth]{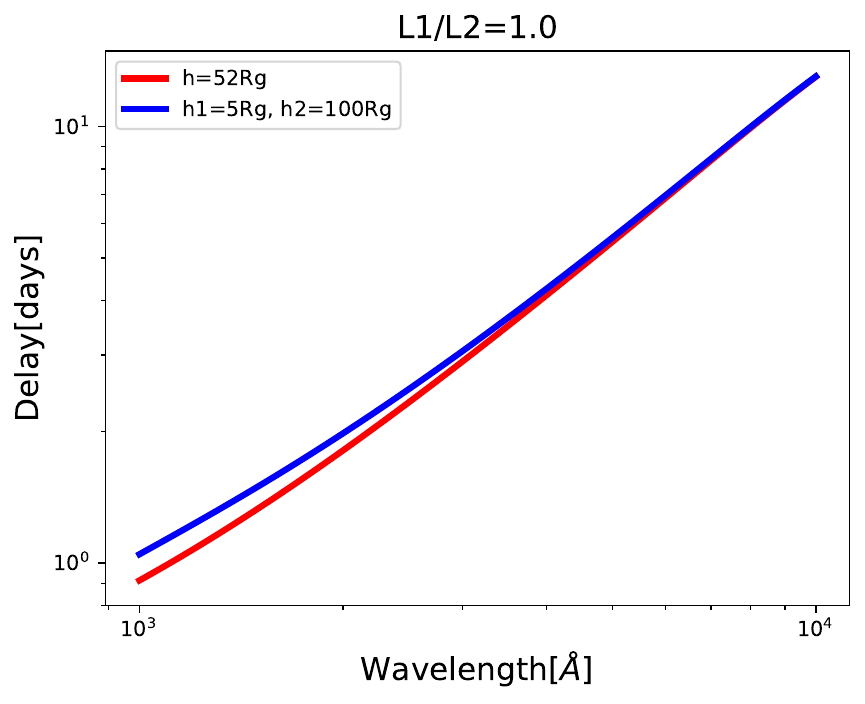}\hfill
            \includegraphics[width=.3\textwidth]{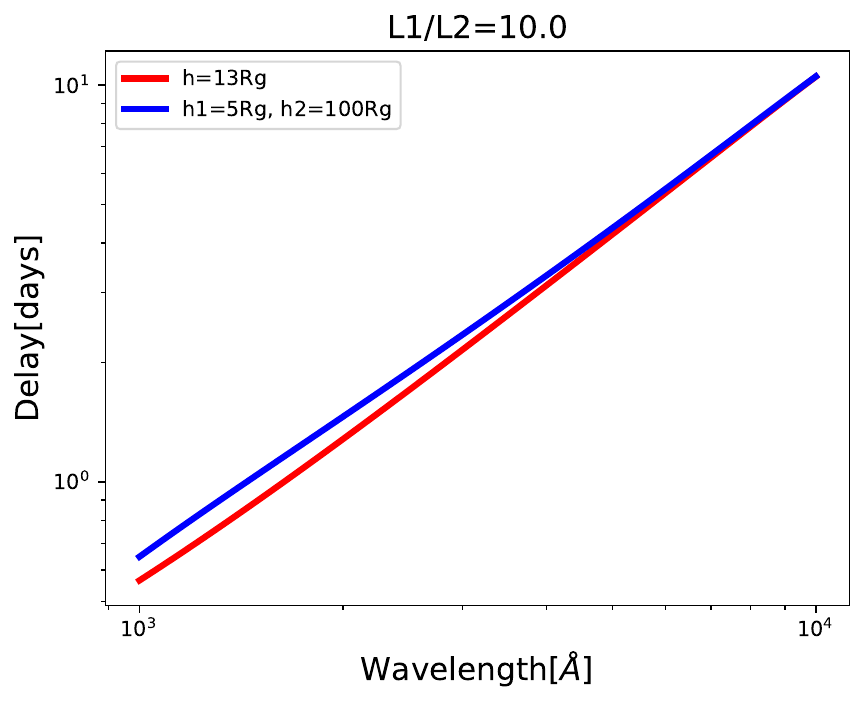}\hfill
            \includegraphics[width=.3\textwidth]{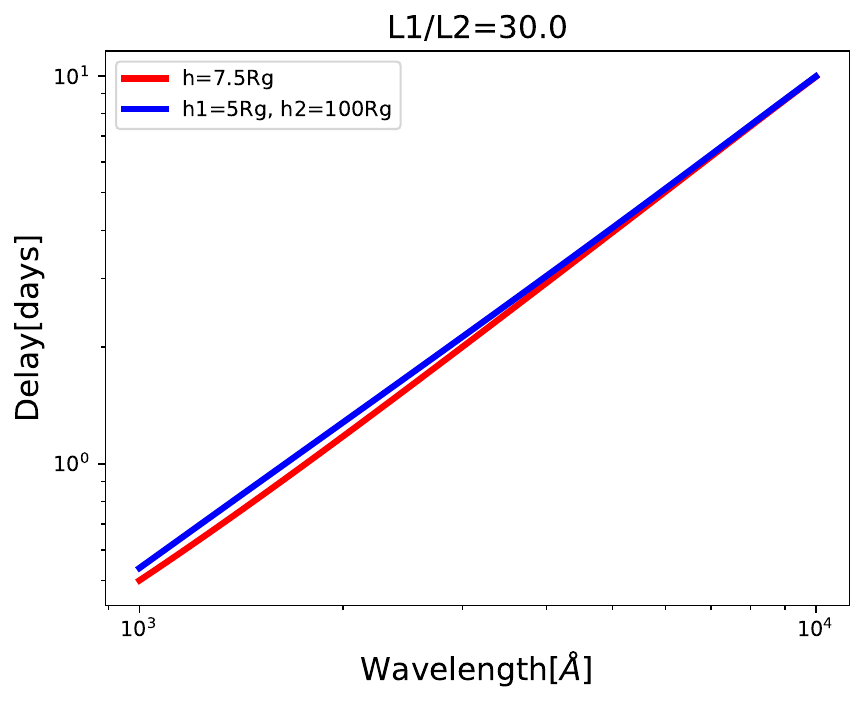}\hfill
            \includegraphics[width=.3\textwidth]{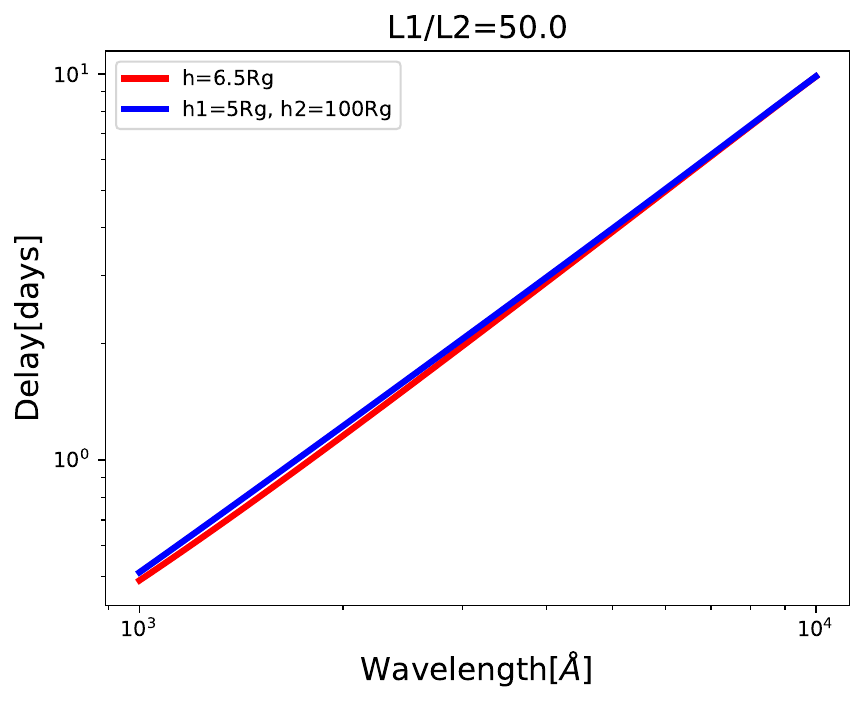}\hfill
            \caption{The plots show the variation in the delay as the luminosity ratio between the two corona is changed. The luminosity for the single corona is fixed at $2 \times 10^{46}$ erg/s, while the height is adjusted such that the delay for the longest wavelength matches the delay for the two corona. The delay from the single corona is represented in red, while the delay from the two corona is shown in blue for different luminosity ratios.
}
\label{fig:delay_comp}
\end{figure*}

\begin{figure}
\centering
\includegraphics[width=0.5\textwidth]{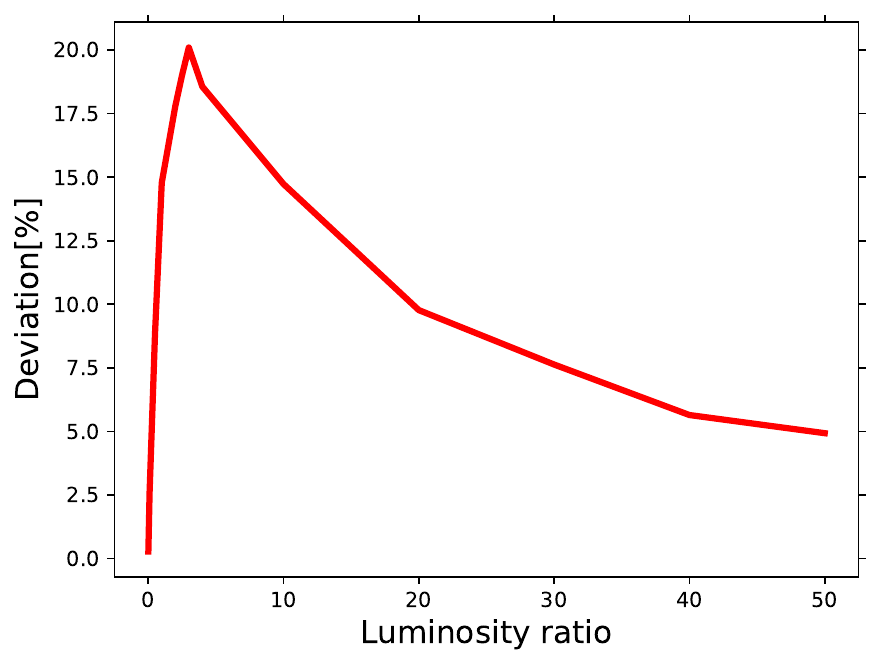}
\caption{Deviation in delay for wavelength 1000$\AA$ for different luminosity ratios as shown in table \ref{tab:max_dev}.}
 \label{fig:dev_del}
\end{figure}

We perfomed the computations for a range of luminosity ratios between the lower corona's luminosity to the upper corona's luminosity. 
The results are shown in Figure \ref{fig:delay_comp}, the blue line represents the delay profile of 2-corona while the red line represents the delay profile of a single corona with combined luminosity and the height optimized as described above. Then, we determined the maximum deviation in delay between the two plots for a given wavelength as shown in Figure~\ref{fig:dev_del}. The results are summarized in Table \ref{tab:max_dev}.

\begin{center}
\begin{table}[]
    \centering
    \begin{tabular}{||c c c ||}
    \hline
 Luminosity Ratio & Deviation[days] &Deviation[$\%$]\\ [0.5ex] 
 \hline\hline
  0.01&0.0035&0.276\\
 \hline
  0.10&0.0313&2.593\\
 \hline
  0.5&0.0922&8.826\\
 \hline
 1.0&0.1351&14.78\\
 \hline
 2.0&0.1407&17.79\\
 \hline
 2.5&0.1422&19.01\\
 \hline
 3.0&0.1432&20.10\\
 \hline
 4.0&0.1254&18.56\\
 \hline
 10.0&0.0832&14.73\\
 \hline
 20.0&0.0507&9.77\\
 \hline
 30.0&0.0382&7.63\\
 \hline
 40.0&0.0279&5.64\\
\hline
50.0&0.0241&4.93\\
\hline
 \hline
    \end{tabular}
    \caption{Difference in delay between single and 2-corona for $1000 \AA$ from Figure \ref{fig:delay_comp}.}
    \label{tab:max_dev}
\end{table}
\end{center}

 We see that the maximum deviation from a single lamp model is at the lamp luminosity ratio 3, and it reaches 20 \%.Such a measurement would, of course, require coverage over a wide range of time delays, from $1000 Å$ to $10000 Å$, with no contamination from BLR scattering. However, with excellent data, these tests can be successfully performed. Overall, large deviations are seen for the lamp luminosity ratio from 2 to 4, and departures are expected to be much lower outside this range.

The effective lamp position is well approximated by the arithmetic average of the assumed lamp positions weighted with the lamp luminosity 
\begin{equation}
h = {h_1 L_1 + h_2 L_2 \over L_1 + L_2}.
\label{eq:effective_height}
\end{equation}

\subsection{cross-correlation function from the two lamp model}

In the case of the two-lamp geometry, the response function has a two-peak shape, hence in principle two timescales from two lamps are present in the process. Therefore, we aimed to check if this could be also present in cross-correlation function, and help to differentiate between a point-like and extended corona. 


We conducted simulations to distinguish between the single corona and two corona models. We generated X-ray light curves using the \citet{timmer1995} algorithm. We then used these light curves to create reprocessed light curves from the disk (see Section~\ref{sect:creating_curves}).  

\begin{figure*}[ht!]
            \includegraphics[width=.2\textwidth]{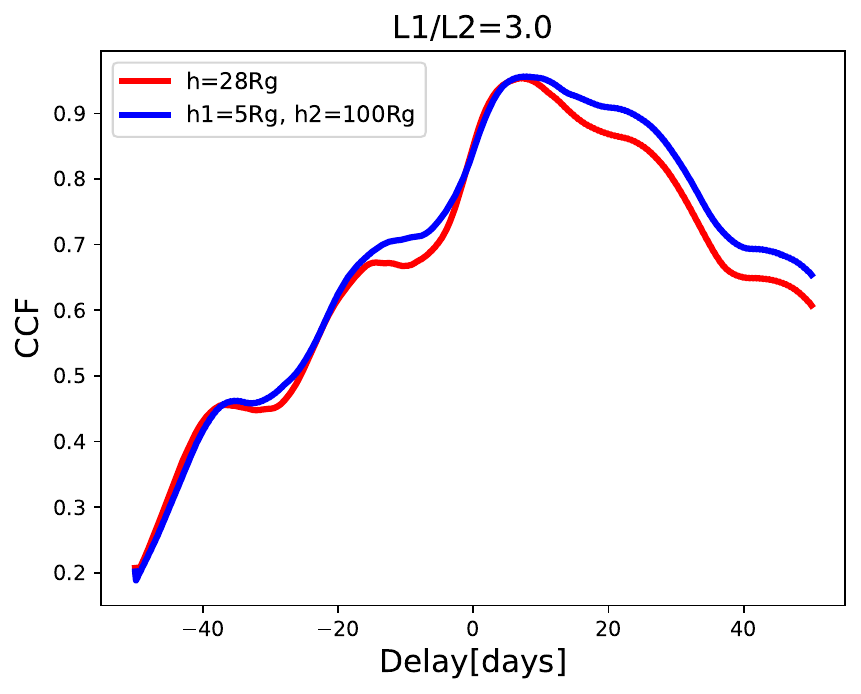}\hfill
            \includegraphics[width=.2\textwidth]{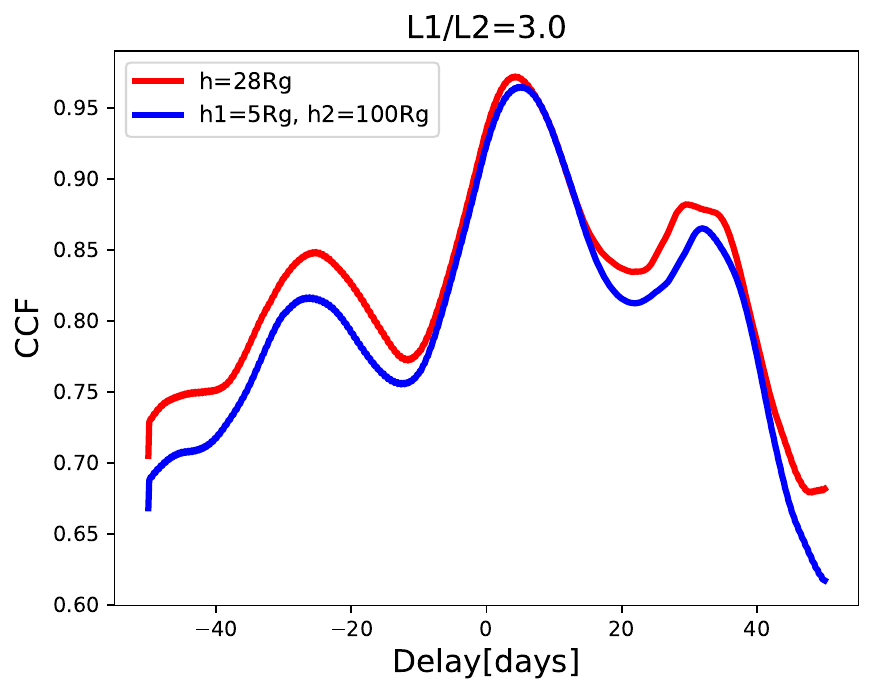}\hfill
            \includegraphics[width=.2\textwidth]{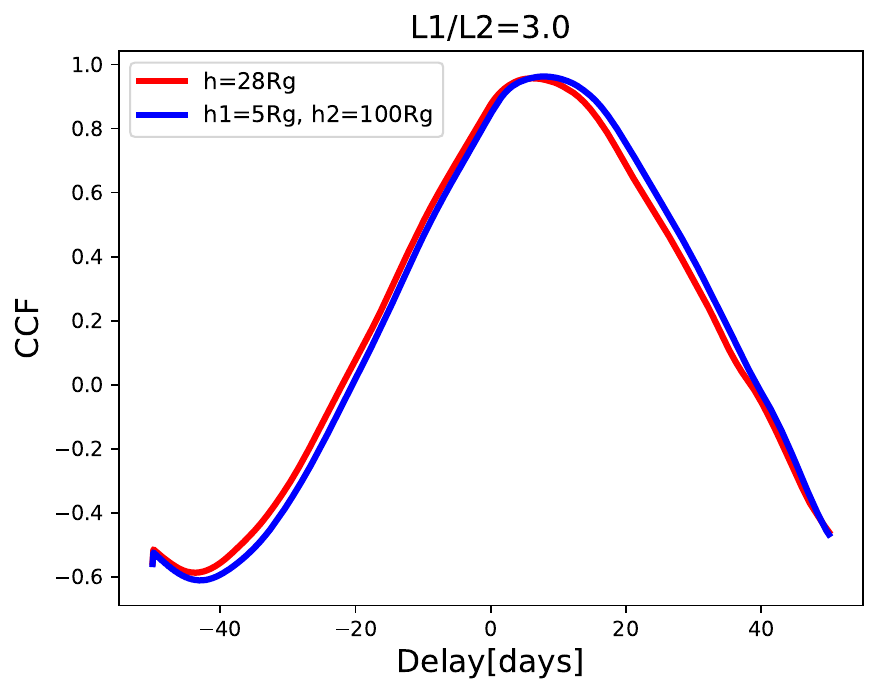}\hfill
            \includegraphics[width=.2\textwidth]{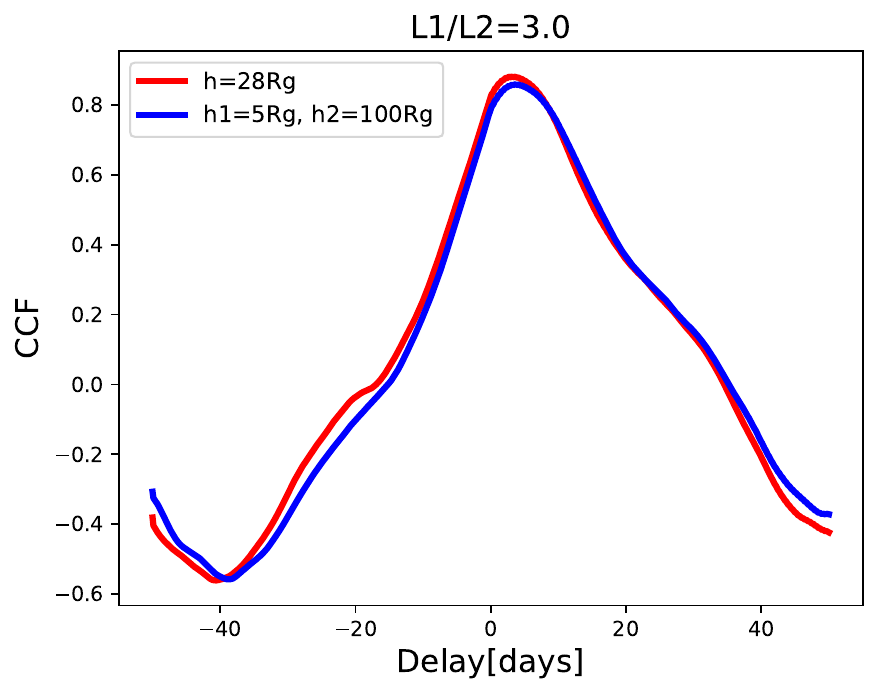}\hfill
            \includegraphics[width=.2\textwidth]{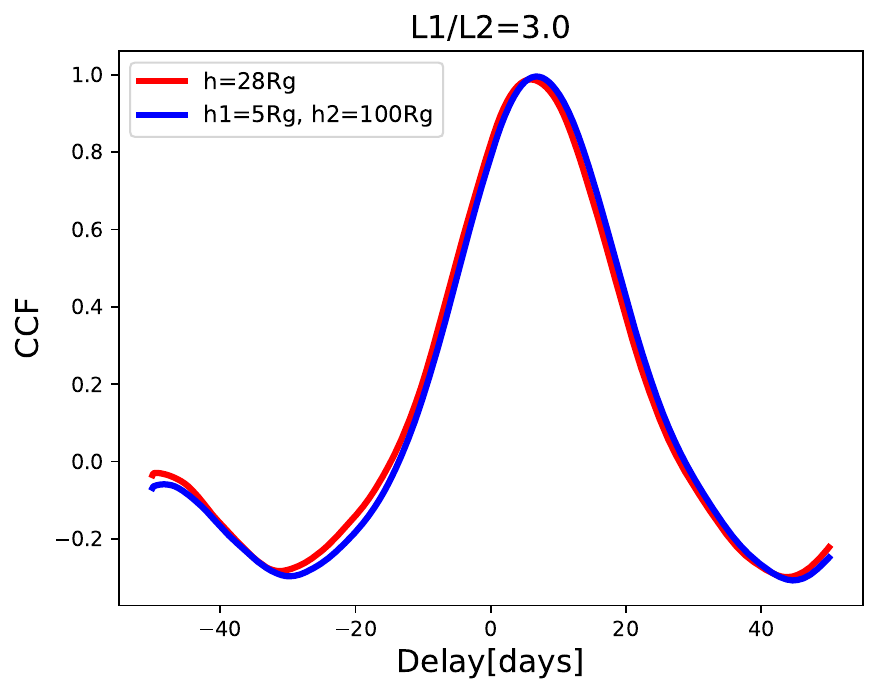}\hfill
            \includegraphics[width=.2\textwidth]{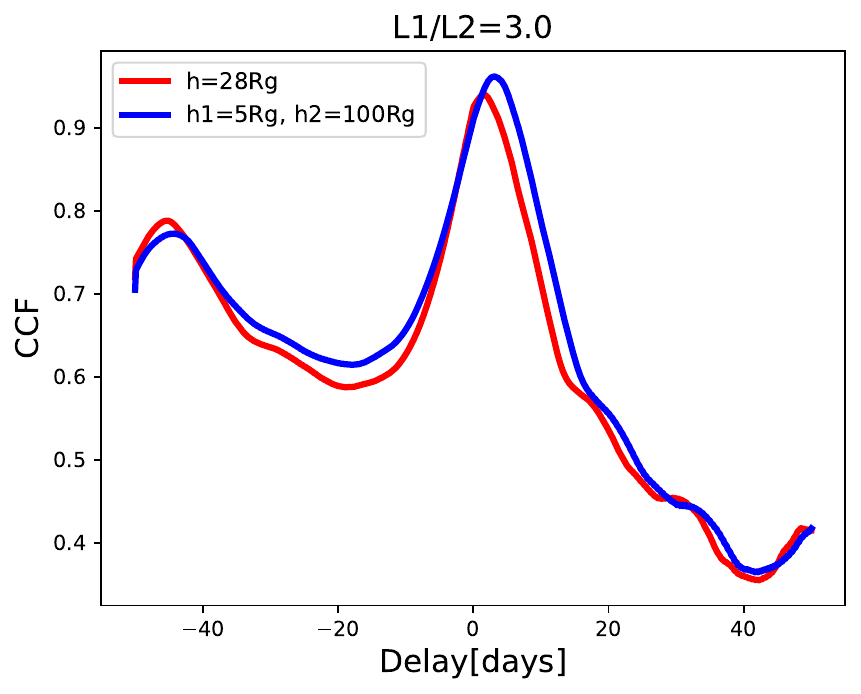}\hfill
            \includegraphics[width=.2\textwidth]{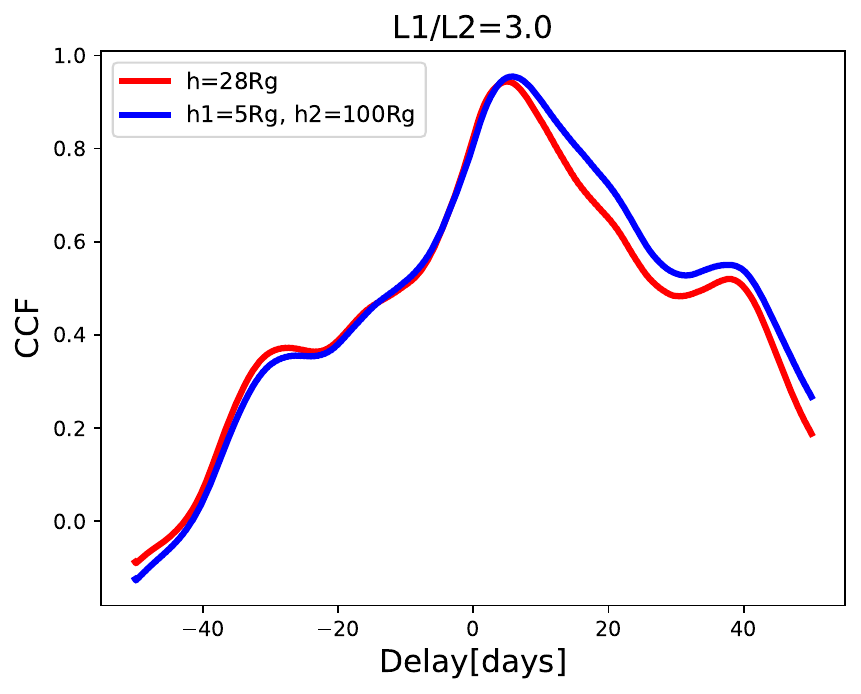}\hfill
            \includegraphics[width=.2\textwidth]{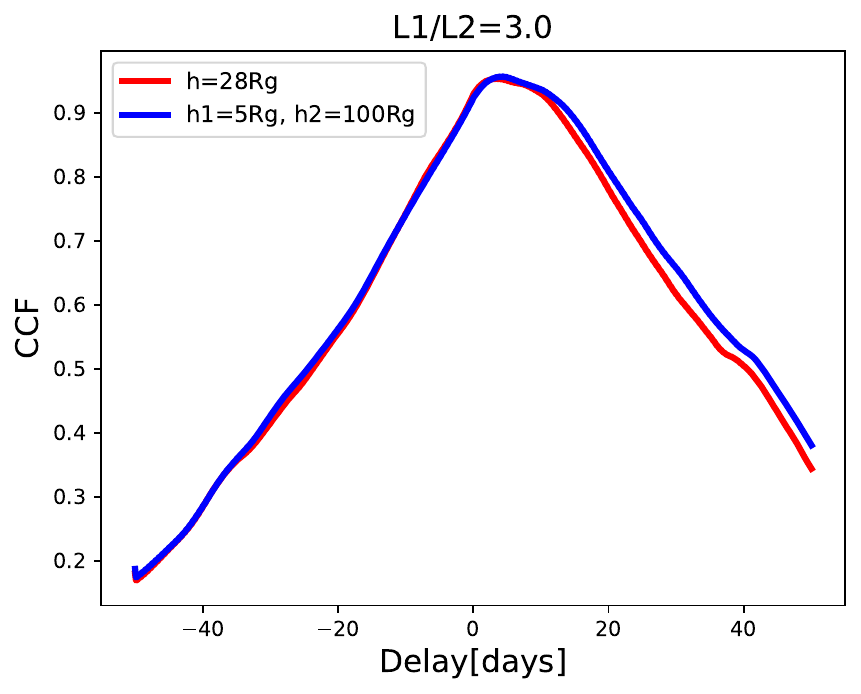}\hfill
            \includegraphics[width=.2\textwidth]{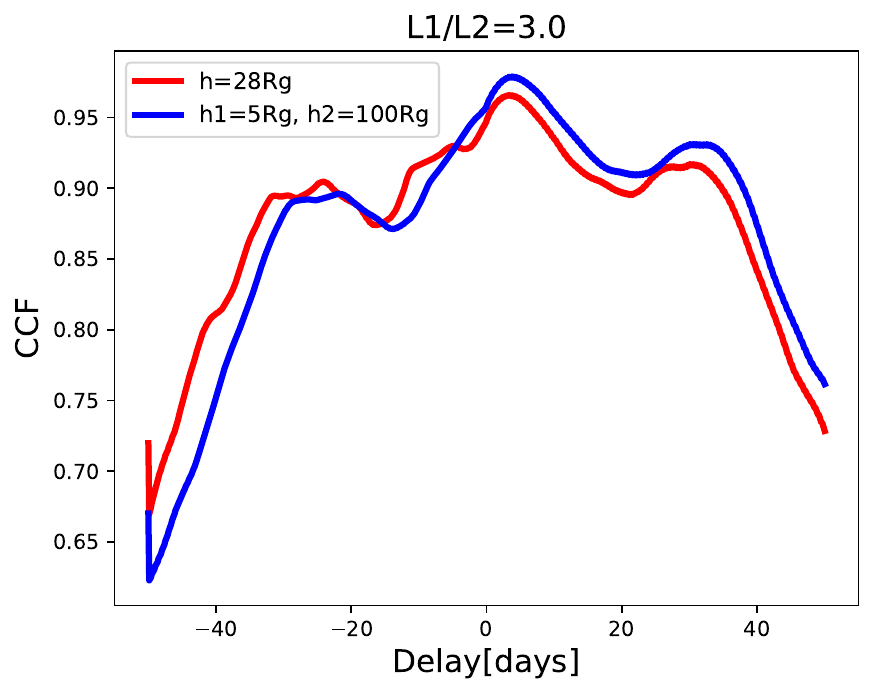}\hfill
            \includegraphics[width=.2\textwidth]{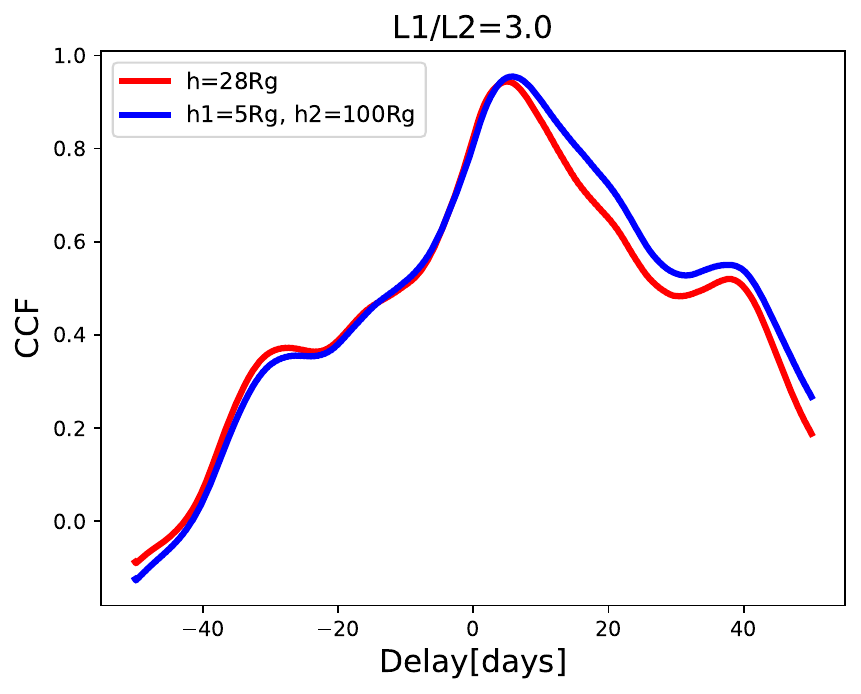}\hfill
            \caption{Examples of the ICCF results for statistically equivalent light curves are shown. The blue line represents the ICCF for the two-corona case, while the red line represents the ICCF for the single-corona case. In this simulation, the luminosity ratio between the two coronae is set to 3. Stochastic variability dominates under the adopted parameters: a light curve duration of 200 days and a sampling interval of 0.1 days.For the single corona, the parameters are: black hole mass $10^{8}M_{\odot}$, incident luminosity $L_X = 2.0 \times 10^{46}$ erg s$^{-1}$, Eddington ratio $= 1.0$, heights $h = 28r_g$, and viewing angle $i = 30^\circ$. For the two coronae, the parameters are: black hole mass $10^{8}M_{\odot}$, luminosities $L_1=1.5\times10^{46}$ and $L_2 = 0.5 \times 10^{46}$ erg s$^{-1}$, Eddington ratio $= 1.0$, heights $h_1 = 5r_g$ and $h_2 = 100r_g$, and viewing angle $i = 30^\circ$.
}
\label{fig:ccf}
\end{figure*}

Next, we calculated the Interpolated Cross-Correlation Function (ICCF) between the two wavelengths, $1000 Å$ and $7923 Å$. This process was repeated for 10 random realizations of the initial light curve, while maintaining the same statistical setup. We used exactly the same lightcurve for a single-lamp corona and for two-lamp model. The 10 ICCF plots are shown in the figure \ref{fig:ccf}. All curves should be statistically equivalent, and the difference shows the expected statistical error for the adopted sampling. The overall shape varies between the different realizations of the process, particularly at the longest timescales. We observe that the ICCF for two corona is consistently broader than that of the single corona, but the differences between individual realizations dominate. 

\begin{figure}
\includegraphics[width=.49\textwidth]
{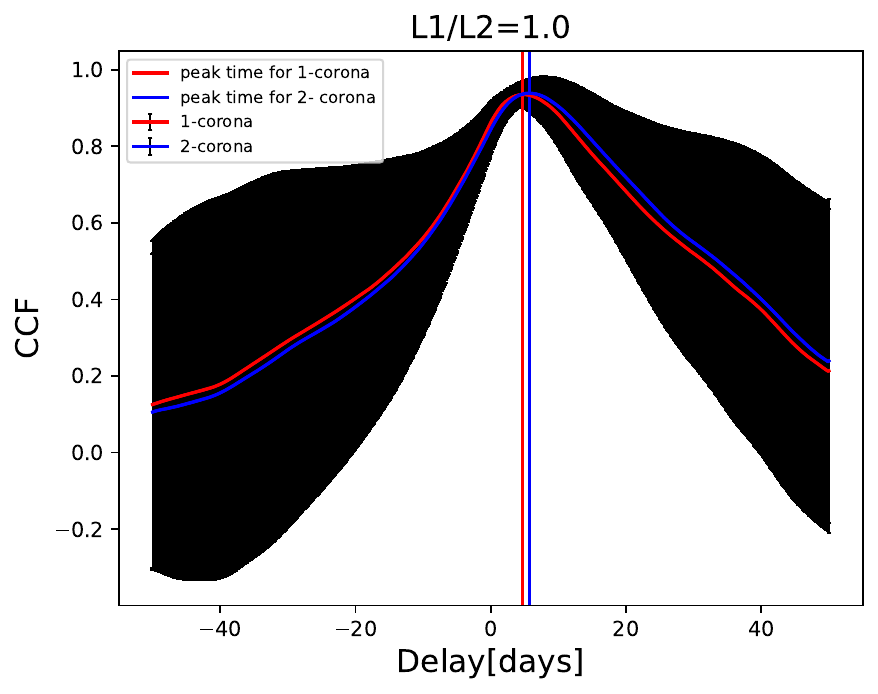}
\includegraphics[width=.49\textwidth]
{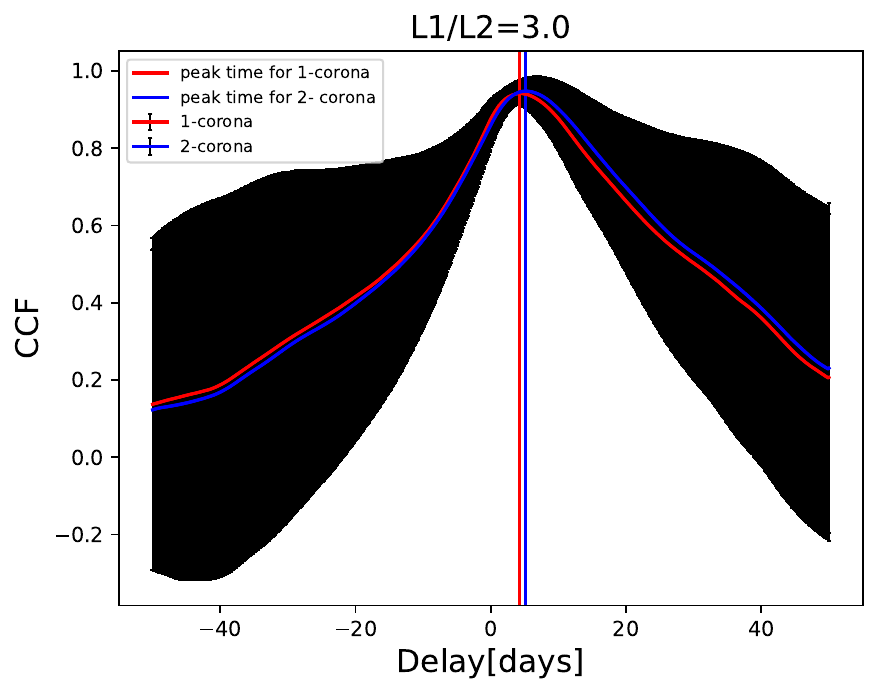}
\caption{Upper panel: The mean and dispersion of 10 statistical realizations of the ICCF for a lamp luminosity ratio of 1:1. Bottom panel: The mean and dispersion of 10 statistical realizations of the ICCF for a lamp luminosity ratio of 1:3. The red line represents the mean ICCF for a 1-corona, while the blue line represents the mean ICCF for 2-corona. The shaded region indicates the dispersion in the 2-corona model.}
\label{fig:ccf_dispersion}
\end{figure}

These 10 exemplary statistical realizations of the lightcurve can be used as a qualitative test whether the two-lamp corona can be differentiated from a single-lamp corona. With this aim, we calculated the mean shape of the ICCF shown in Figure~\ref{fig:ccf}, and the contour plot shows the dispersion. The result for the lamp luminosity ratio 1:1(top panel) and 1:3(bottom panel) are shown in Figure~\ref{fig:ccf_dispersion}. We observe that the two mean curves lie well within the dispersion, making it difficult to distinguish between them based on a single measurement. The error of the mean would be by a factor of $\sqrt{N}$ lower than the dispersion, but even 10 lightcurves for a given source would not allow us to see the difference. Denser monitoring than 10 observation per day is unlikely, but eventually much longer observations (several seasons) could help. 

\subsection{comparison of ICCF time delay with respect to single corona }
In Figure~\ref{fig:ccf_dispersion}, we present the averaged values and standard deviations of all 10 ICCFs for both single and double corona configurations. Simulations were conducted for luminosity ratios of 1 and 3, with the results summarized in Table~\ref{tab:peak_time}.
For a luminosity ratio of 1, the peak occurs at 4.69 days for the single corona configuration and at 5.69 days for the double corona configuration. Similarly, for a luminosity ratio of 3, the peak is observed at 4.19 days for the single corona and at 5.19 days for the double corona. Thus, extended corona increases statistically the mean time delay but fluctuations are large.  
\begin{center}
\begin{table}[]
    \centering
    {\small
    \begin{tabular}{||c c c ||}
    \hline
    Luminosity Ratio & $(t_{peak})_{1-corona}$ [days] & $(t_{peak})_{2-corona}$ [days] \\[0.5ex] 
    \hline\hline
    1.0 & 4.69 & 5.69 \\
    \hline
    3.0 & 4.19 & 5.19 \\
    \hline
    \end{tabular}
    }
    \caption{Table summarizing the peak delay from figure \ref{fig:ccf_dispersion}.}
    \label{tab:peak_time}
\end{table}
\end{center}




\section{Discussion}

We presented a simulated setup for irradiation of the disk by the vertically extended hot corona. We aimed to see if the optical/UV reverberation mapping of the accretion disk can reveal such an geometry. The simulations were performed for a black hole mass $10^8 M_{\odot}$, dense ( 0.1 day sampling) and long (200 days) monitoring. The expected delays for such setup were from fraction of a day at short wavelengths to 10 days at $\sim 10000$ \AA. The location of the two lamps were extreme, at $5 r_g$ and $100 r_g$. When one of the lamps dominates, the wavelength-dependent delay pattern is like for a single lamppost model. When the two sources have comparable luminosity there is a deviation between the two-source pattern and a single-source (intermediate height) pattern but smaller than 20 \%. Therefore, it is not very likely to resolve the vertical extension of the hot corona with the current studies of the continuum time delays. The eventual traces of the vertical extension of the corona in the shape of ICCF are even harder to detect under the discussed setup. 

The two lamps are roughly equivalent to a single lamppost model with the location determined as the luminosity-weighted average height (see Equation~\ref{eq:effective_height}). This is simply related to the fact that most reprocessing observed in optical/UV happens at a distance larger than the largest corona height. In this case the local ratio of the incident radiation from all N lamps to the disk emission is independent from the radius, and given by the ratio
\begin{equation}
{F_{inc} \over F_{disk}} \propto \sum_{i=1}^N L_i H_i,
\end{equation}
as implied by the ratio of the lamp terms to the first term corresponding to non-illuminated disk in Equation~\ref{eq:eq6}. 
Similar condition holds for a continuum distribution, then the sum would be replaced with the integral. 

When the two lamps have the same luminosity the effective height is practically just half of the higher lamp position. Therefore, the upper extension of the corona in this sense dominates. This is due to the simple fact, that the irradiation due to the higher lamp dominates at distances larger than $\sim 30 r_g$, as illustrated in Figure~\ref{fig:schematic} where the emission at longer wavelengths is produced. 

\begin{figure}
\includegraphics[width=.5\textwidth]{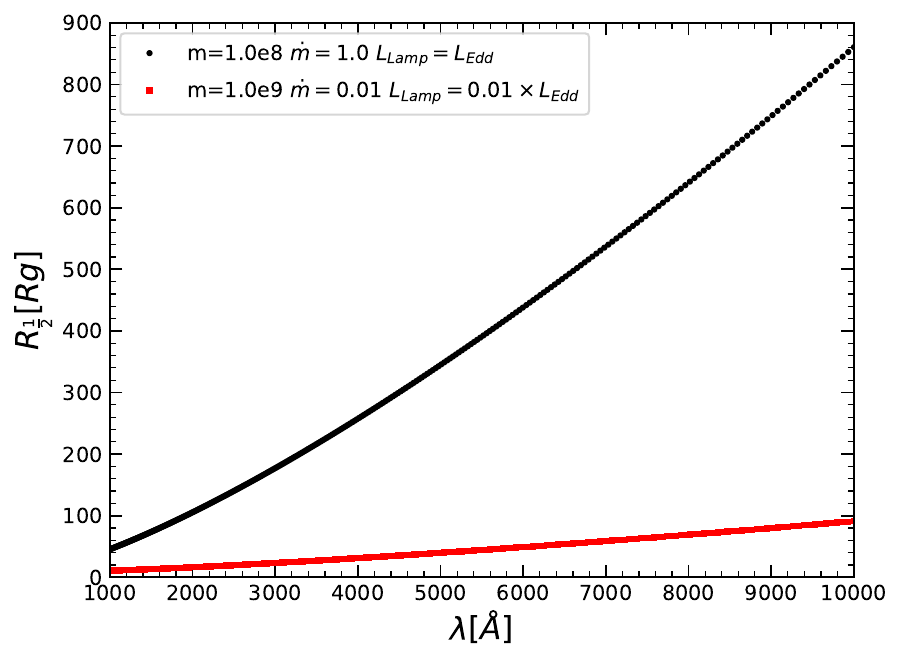}
\caption{The dependence of the mid-radius, where 50\% of the radiation is emitted, on the emission wavelength is shown for two models: one with a black hole mass of $10^8 M_{\odot}$ and an Eddington rate of $\dot{m} = 1$ (black line), and another with a black hole mass of $10^9 M_{\odot}$ and an Eddington rate of $\dot{m} = 0.01$ (red line).
 }
\label{fig:mid_radius}
\end{figure}

In our standard model the emission above 2000 \AA~ originates at a distance larger than the maximum height of the corona, 100 $r_g$ , as illustrated by Figure~\ref{fig:mid_radius}. However, this value depends on the black hole mass as well as on the Eddington ratio. We may have better prospects to see the effect of extended corona not equivalent to a mean position when the Eddington ratio is lower, and the black hole mass is higher. This is illustrated in Figure~\ref{fig:mid_radius} with the red line. In this case the reprocessing takes place at the radii lower or comparable to the height of the upper corona. This opens a prospect to see the difference between a single lamppost and an extended corona. We check that showing the expected time delay.

\begin{figure}
\includegraphics[width=.5\textwidth]{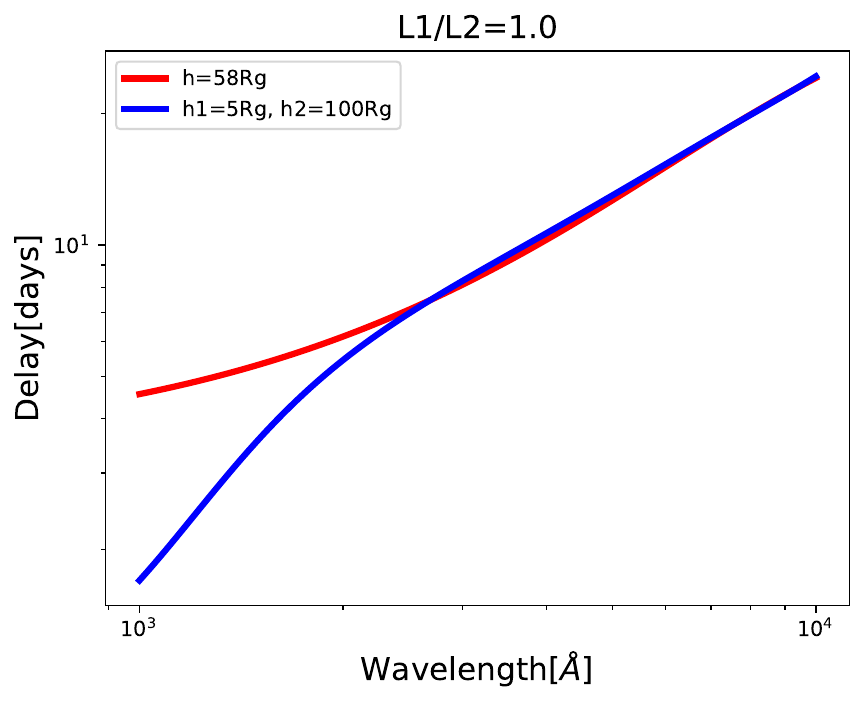}
\caption{The plot compare the delays for a 1-corona and 2-corona (with a luminosity ratio of 1). The height of the single corona is selected such that the delay for the longest wavelength matches that of the 2-corona model. For the single corona, the parameters are: black hole mass $10^{9}M_{\odot}$, incident luminosity $L_X = 2.0 \times 10^{46}$ erg s$^{-1}$, Eddington ratio $= 0.01$, height $h = 58r_g$, and viewing angle $i = 30^\circ$. For the two coronae, the parameters are: black hole mass $10^{9}M_{\odot}$, luminosities $L_1 = 1.0 \times 10^{46}$ erg s$^{-1}$ and $L_2 = 1.0 \times 10^{46}$ erg s$^{-1}$, Eddington ratio $= 0.01$, heights $h_1 = 5r_g$ and $h_2 = 100r_g$, and viewing angle $i = 30^\circ$.
}
\label{fig:delay_for_e9}
\end{figure}

The delay from the two-corona model in the case of the large black hole mass and low Eddington rate is shown in Figure~\ref{fig:delay_for_e9}. The delay now does not follow the standard $\lambda^{4/3} $ law any more, as it bends rapidly at the shortest wavelengths. Due to the irradiation, the temperature is almost constant at small radii where radiation at the shortest wavelengths is produces (see the schematic picture in the lower panel of Figure~\ref{fig:schematic}).  Therefore, the time delay almost disappears.
However, most of the reverberation-measured sources do not have such high masses and low Eddington ratios. In addition, our assumptions underlying the model may not be satisfied in this case. We assume (see Section~\ref{sect:disk}) that the standard disk extends down to ISCO while this is not true for lower values of the Eddington ratio, and the transition to an inner hot flow is not well described. In addition, since in such case all reprocessing takes place closer to the black hole, the effects of General Relativity become important, and the use of geometrical optics is not justified. If there is a suitable observational data then much more advanced modeling would have to be performed, and the methodology of \citet{langis2024} would be an excellent starting point. 


\section{Conclusions}

\begin{itemize}
    \item \textbf{Sensitivity of UV/Optical Reverberation Mapping:} The study reveals that UV/optical reverberation mapping has limited sensitivity to the vertical extension of the corona. Detecting such an extension requires highly precise time delay measurements (with errors below 20\%) and exceptionally dense sampling of lightcurves.
    
    \item \textbf{Effective Position of the Corona:} The effective position of the corona is well approximated by a luminosity-weighted average height of the emitting components. This result underscores that the impact of vertical extension on reverberation mapping is minimal under typical observational conditions.
    
    \item \textbf{Comparison of Two-Lamp and Single-Lamp Models:} The differences in wavelength-dependent delays between the two-lamp corona model and the single-lamp model are relatively small, with a maximum deviation of 20\%, observed for a lamp luminosity ratio of about 3. These differences remain challenging to detect with current observational techniques.
    
    \item \textbf{Insights from Cross-Correlation Function (ICCF) Analysis:} Simulations demonstrate that ICCFs for single-lamp and two-lamp corona configurations are statistically similar. Differentiating the two geometries would require significantly extended monitoring periods or much denser observational sampling.
    
    \item \textbf{Role of Black Hole Mass and Eddington Ratio:} The study suggests that larger black hole masses and lower Eddington ratios enhance the likelihood of detecting the effects of an extended corona. These conditions result in more pronounced deviations from single-lamp model predictions, but would require much more advanced modelling.
    
\end{itemize}
\section*{Acknowledgments}
We are thankful to Amit Kumar Mandal for helpful suggestions in the text. This project has received funding from the European Research Council (ERC) under the European Union’s Horizon 2020 research and innovation program (grant agreement No. [951549]). VKJ acknowledges the OPUS-LAP/GAČR-LA bilateral project (2021/43/I/ST9/01352/OPUS 22 and GF23-04053L).

\bibliographystyle{aa}
\bibliography{main}

\end{document}